\documentclass[ superscriptaddress,showpacs,twocolumn,amssym]{revtex4-1}
\usepackage{latexsym}
\usepackage{graphicx}
\usepackage{rotating}
\usepackage{longtable}
\usepackage[usenames]{color}
\usepackage[normalem]{ulem}

\usepackage{multirow} 

\input colordvi

\begin{document}

\title{Optimal interface doping at { $\rm \bf    La_{2/3}Sr_{1/3}MnO_3/SrTiO_3(001)$} 
heterojunctions for spintronic applications}

\author{C. Wang}
\affiliation{Abdus Salam International Centre for Theoretical Physics, 
Strada Costiera 11, Trieste 34151, Italy}

\author{N.~Stoji\'{c}}
\affiliation{Abdus Salam International Centre for Theoretical Physics, 
Strada Costiera 11, Trieste 34151, Italy}
\affiliation{ IOM-CNR Democritos,  Trieste, I-34151, Italy}

\author{N.~Binggeli}
\affiliation{Abdus Salam International Centre for Theoretical Physics, 
Strada Costiera 11, Trieste 34151, Italy}
\affiliation{ IOM-CNR Democritos,  Trieste, I-34151, Italy}

\date{\today}

\begin{abstract}

We examine, by means of {\it ab~initio} pseudopotential calculations, 
$\rm La_{2/3}Sr_{1/3}MnO_3 / SrTiO_3$  (LSMO/STO) heterojunctions in which
one unit layer of   $\rm La_{1-x}Sr_{x}MnO_3$ (with $\rm 0<x<1$) is
inserted at the interface. The optimal interlayer doping 
x for a robust interface ferromagnetism is investigated by considering
the  energy differences between
antiferromagnetic and ferromagnetic alignment of the $\rm
MnO_2$-interface layer relative to bulk LSMO. 
The optimal doping is found to be close to $x=1/3$, 
which corresponds to an abrupt $\rm TiO_2$ (001)-layer termination of STO. This
is also the composition which gives the largest $p$-type Schottky
barrier height in our calculations.

\end{abstract}


\maketitle

Very promising magnetic tunneling junctions (MTJ's) based on 
$\rm La_{2/3}Sr_{1/3}MnO_3$ (LSMO) and $\rm SrTiO_3$ (STO)
LSMO/STO/LSMO(001)\cite{KouSonHwa10,YamOgaIsh04} 
have not yet realized their full potential. Due to its large spin
polarization (half or close to half metallicity) and high Curie
temperature ($T_c=370$~K), LSMO is well suited for spintronic applications,  
such as tunneling magnetoresistance (TMR) devices and field-effect
transistors~\cite{MiaMunMoo11, YajHikHwa11,KouSonHwa10,BibBer07}. At the same time,
STO is a wide-gap semiconductor closely
lattice matched to  LSMO (less than 1\%  mismatch) and is therefore
also well suited as tunneling barrier in LSMO/STO/LSMO MTJ's. In fact, 
a TMR ratio as large as $\sim$1900~\% has been measured in LSMO/STO/LSMO MTJ's
at 4~K~\cite{note_papers}.

However, the TMR ratio at higher temperatures drops significantly and
vanishes well below room
temperature~\cite{note_papers,GarBibBar04,SunGalDun96}.  Origins of
this behavior might lie in the presence of extrinsic 
defects, loss of stoichiometry, interface roughness, atomic
intermixing at the
interface~\cite{GraPapBal10,HerWilSch08,SamImhMau03,PaiImhSik02,VerDavMir12}  
 and canting of interface spins~\cite{IzuOgiOki01,OgiIzuSaw03}.

Possibly, such problems could be countered by  atomic-scale control of the
interface properties in epitaxial growth of the heterostructures. In
a pioneering study, Yamada et al.~\cite{YamOgaIsh04,IshYamSat06},
successfully enhanced the ferromagnetism and TMR (from 50~\% to 170~\%
at 10~K) of their  LSMO/STO/LSMO structures; this was achieved by engineering, at the
atomic scale, the interface doping profile with the growth of one
bilayer of $\rm LaMnO_3$ at the nominally SrO-terminated
LSMO/STO(001) interface. However, it should be noted that their TMR
values,  based on the engineered SrO-terminated LSMO/STO(001) 
interface,  were significantly lower than other reported
results on unmodified LSMO/STO/LSMO(001)
structures~\cite{note_papers,GarBibBar04}. More recently, Kourkoutis
et al.~\cite{KouSonHwa10} also showed that
 enhanced ability to control the microscopic growth and interface
 sharpness in LSMO/STO(001) superlattices leads to a large improvement
 $-$with the stabilization at room temperature of 
  ferromagnetism and  metallicity for LSMO thicknesses down to 2~nm~\cite{KouSonHwa10}.

Nonetheless, despite impressive progress achieved so far in the epitaxial control
of the LSMO/STO interface and  atomic-layer engineering of the same, some
uncertainties remain in general  on the details of the atomic
structure. This concerns, in particular, 
the precise chemical stoichiometry of the interface layers, given the significant  atomic
intermixing typically present at such
interfaces~\cite{HerWilSch08,PaiImhSik02,GraPapBal10,BosVerEgo12}.  
This, in fact, hinders a precise systematic assessment of the optimal interface
atomic configuration for a robust ferromagnetism $-$and possibly enhanced TMR$-$  directly
from experiment. Thus, a theoretical study of optimal doping remains
the most direct answer to this question. 
 
Theoretically, the two distinct chemically abrupt SrO- and $\rm
TiO_2$-layer terminations of the STO(001) at the interface are
commonly assumed~\cite{ZenGehTem07,ZheBin10,BurTsy10}. A
strengthening 
of the ferromagnetism at the LSMO/STO interface with the SrO-layer termination upon doping 
with one or two $\rm LaMnO_3$ unit layers inserted at the interface
was obtained performing self-interaction-corrected local spin 
density calculations~\cite{ZenGehTem07}. No $\it ab~initio$
calculation, however, has been carried out yet on the influence of
layer doping on the ferromagnetism of the $\rm TiO_2$-terminated
interface. Only a model inferred from previous
calculations~\cite{ZenGehTem07,ZheBin10}, based on nominal charges for
the Sr, La, Ti, and O ions, may be used to predict the optimal doping
in this case~\cite{note_model}.  Within this model, each 
 $\rm La_{1-x}Sr_xO$ layer distributes (1-x) electrons over the 
adjacent   $\rm MnO_2$ layer(s)~\cite{ZenGehTem07,ZheBin10}.
 For the $\rm
TiO_2$-layer termination, the model would
predict an optimal doping of $\rm x\approx 2/3$ for a $\rm
La_{1-x}Sr_xMnO_3$ unit layer inserted at the LSMO/STO interface. In
this context,
$\it ab~initio$ calculations as a function of x, in the full range
$\rm 0<x<1$ are needed to provide a more precise and reliable
determination of the optimal doping.
A recent first-principles study~\cite{BurTsy10} of the
band alignment at this interface, with x in the range  $\rm 0.5<x<1$,
demonstrated a strong dependence on  x, 
indicating a possibility of controlling the Schottky barrier height (SBH)
by changing the composition of the interfacial $\rm La_{1-x} Sr_xMnO_3$
layer.  The SBH is the fundamental interface parameter that controls the transport
properties of the fully developed metal/semiconductor junction. It can
also be related to the tunneling barrier in MTJ~\cite{BowBarBel06},
although it is only an estimate in this case. The trends with x,
however, are expected to be the same in the two cases.

In this work, we apply the first-principles density-functional theory (DFT)  to systematically
explore effects of the interface layer 
doping, in the whole doping range, on the robustness of the
ferromagnetism at the LSMO/STO(001) 
heterojunction. Our findings indicate an optimal doping close to $\rm
x=1/3$. In
addition, we calculate the $p$-type SBH for all the dopings considered
and find that it also has the maximum value for $\rm x\approx1/3$.



Our DFT calculations were performed using the PWSCF code~\cite{GiaBarBon09}
with ultrasoft pseudopotentials~\cite{Van90,pseudo} and plane
wave basis set. For exchange and correlation, we used  the generalized gradient
approximation (GGA) in the Perdew-Burke-Ernzerhof
parametrization~\cite{PerBurErn96}. In addition, calculations on all configurations
were also performed using the GGA plus on-site Coulomb interaction
approach (GGA+U)~\cite{AniAruLic97}.
We applied a kinetic-energy cutoff of 30 Ry for
the plane-wave expansion of the electronic wavefunctions and of 350
Ry for the electronic charge density.  The Brillouin-zone sampling was
performed using a  $6\times2\times1$ k-point grid centered at
$\Gamma$. We employed a Gaussian-level smearing of 0.01~Ry to
determine the Fermi energy. 

\begin{figure}[tp]
\begin{center}
\includegraphics[width=2.5cm]{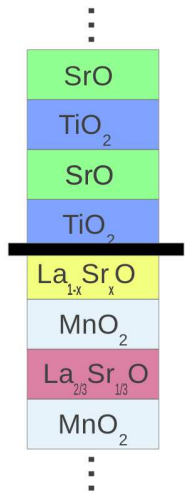}
\caption{ A schematic representation of the interface layer
  doping considered in this work. The  $\rm La_{2/3}Sr_{1/3}O$ layer at the interface  is
  replaced by $\rm  La_{1-x}Sr_{x}MnO_3$. Different layers in STO and
  LSMO are color-coded, with the   interface (denoted with a black
  line) located between  $\rm La_{1-x}Sr_xO$ and $\rm   TiO_2$ layers
  ($\rm TiO_2$ termination of STO).  }   
\label{fig:doping}
\end{center}
\end{figure}

\begin{figure}[htp]
\begin{center}
\includegraphics[width=8.5cm]{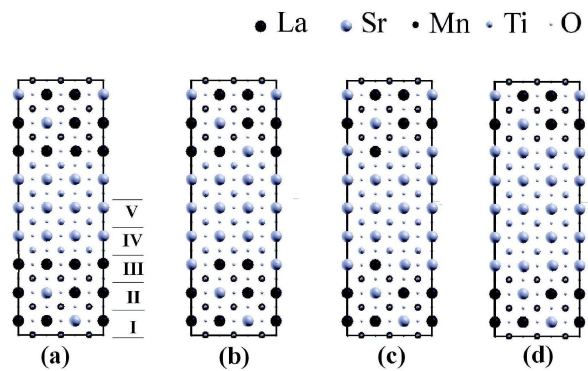}
\caption{Supercells (periodically repeated) used to model the LSMO/STO(001)
  heterojunctions with different interface doping configurations. The
  regions of the supercell labeled I to V include each two monolayers
  forming an LSMO
  unit (regions I to III) or an STO unit (regions IV and V). The
  doping of the   interface  $\rm La_{1-x}Sr_{x}MnO_3$ layer, (in region
  III) takes on the values:  $ x= 0$ (a), $1/3$ (b), $2/3$ (c) and $1$
  (d). Large black, large grey, small black and small grey spheres
  denote La, Sr, Mn, and Ti atoms, respectively. The grey dots
  indicate the O atoms. }
\label{fig:model}
\end{center}
\end{figure}

In our study, the $\rm La_{2/3}Sr_{1/3}$ alloy in LSMO is modeled using
an ordered alloy configuration with an identical stoichiometry
of  La and Sr atoms in each LSMO(001) layer~\cite{ZheBin09}, as
opposed to using the virtual crystal 
approximation~\cite{ZenGehTem07, BurTsy10}. The doping is realized by
replacing the $\rm La_{2/3}Sr_{1/3}O$ layer at the interface by a layer with a variable
doping of the form $\rm La_{1-x}Sr_{x}O$ (with $\rm x=1$, $2/3$, $1/3$ or $0$), as shown in
Fig~\ref{fig:doping}. The STO(001) slab is terminated with the $\rm TiO_2$
layer. The case with  $\rm x=1/3$ doping corresponds to the abrupt $\rm
TiO_2$-termination, while the interface with doping $\rm x=1$ can be
viewed as the abrupt STO(001) $\rm SrO$-terminated interface shifted one layer deeper.
The supercells we employed to model the LSMO/STO(001) heterojunctions with
different interface doping configurations are shown in
Fig.~\ref{fig:model}. They contain each 135 atoms, corresponding to 9
perovskite-layer units along the [001] growth direction and to a
lateral dimension of $1\times3$ perovskite-surface units. Each
supercell includes two equivalent interfaces
between the LSMO and STO regions. The bulk region of LSMO has a
homogeneous Sr-bulk-doped configuration with identical $\rm
La_{2/3}Sr_{1/3}MnO_3$ stoichiometry in each LSMO(001)
layer~\cite{ZheBin09}. To
construct the supercells, we have initially assumed a structure with
ideal perovskite
units and used the perovskite lattice parameter ${a = 3.88}$~\AA,
taken from our previous GGA study of the bulk properties of $\rm
La_{2/3}Sr_{1/3}MnO_3$~\cite{ZheBin09}. This value is very close to the
experimental value of the LSMO lattice parameter
(3.87~\AA)~\cite{TsuSmoNat00} and only slightly smaller (less than 1~\%
smaller) than that of STO (3.905~\AA)~\cite{WoiLiZsc06}. The tilting of
the $\rm MnO_6$ octahedra \cite{GenZha06} has been neglected.
 The effect of atomic
relaxation at the interface was then included by optimizing the
positions of the atoms of the two outermost monolayers of LSMO and of
STO at the junctions (region III-IV and equivalent in Fig.~\ref{fig:model})
 until the residual forces on these atoms were smaller than
$10^{-3}$~Ry/bohr.

As shown previously~\cite{ZheBin10}, GGA  largely underestimates
the STO bandgap (1.9~eV versus the experimental
3.25~eV~\cite{BenElsFre01}). Depending on the STO terminations, this can result
in  $p$-type Schottky barrier heights exceeding the STO
bandgap. This, in turn, causes an artificial transfer of electrons from the
metal to the STO which makes the calculated $p$-type SBH value
saturate at about the bandgap value. We have thus used in the present
work also the  GGA+U which corrects the STO
gap~\cite{ZheBin10}.   The values we use for the Hubbard U's acting on 
the Ti and Mn 3d states,  $\rm U_{Ti}=8~eV$ and $\rm
U_{Mn}=2~eV$~\cite{ZheBin10}, reproduce  the experimental
STO bandgap and LSMO half metallicity.

\begin{figure}[t]
\begin{center}
\includegraphics[width=8.3cm,angle=0]{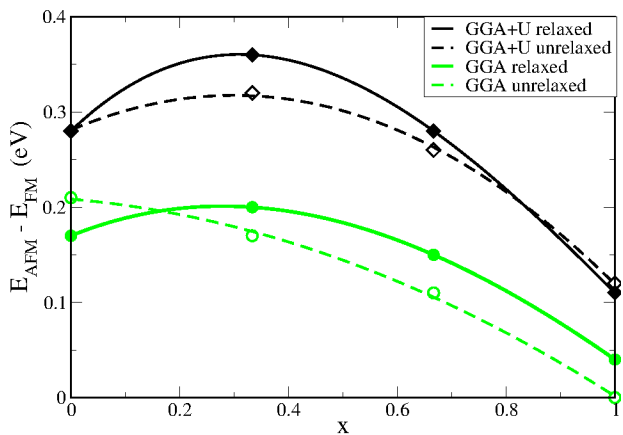}
\caption{ Exchange energy for $\rm LSMO/STO(001)$ heterojunctions with
  different doping levels x. The solid  and dashed lines are cubic spline interpolations of the
  data sets. The energies are given in electron volt per LSMO/STO
  ($1\times3$) interface unit. The curves denoted as ``unrelaxed'' are
obtained for the ideal cubic perovskite atomic structure, without optimization of the
atomic positions at the interface.}
\label{fig:energy}
\end{center}
\end{figure}

As a measure of robustness of the interface ferromagnetism, 
we have evaluated the exchange energy~\cite{ZheBin10},
$E_{AFM}-E_{FM}$, {\it i.e.} the
energy difference between antiferromagnetic and ferromagnetic
alignment of the single $\rm MnO_2$-interface layer relative to bulk
LSMO  (we always assume  a ferromagnetic spin order within each  
 $\rm MnO_2$ plane).  
The calculated  exchange energies are shown in Fig.~\ref{fig:energy}.
 The results are reported for
different interface doping configurations.  Both GGA+U and GGA results
are included for comparison. We first analyze the curves obtained for
the structures in which atomic positions at the interface were optimized (''relaxed'').
Our calculations reveal that the FM alignment of Mn moments is lower
in energy than the AFM alignment for all  considered junctions, as 
shown in Fig.~\ref{fig:energy}. The optimal doping, estimated as  the maximum of the
spline curves in Fig.~\ref{fig:energy},  is rather close  to $x=1/3$. 
Overall, the exchange energy is
strongly influenced by the doping level. 
At $x= 1$, the exchange energy  is the smallest. It amounts to 13 and
37~meV per Mn interface atom in the GGA and GGA+U calculations,
respectively, which is comparable to the room temperature kT. With the
decrease of x, the exchange energy increases steadily until $\rm x=1/3$,
where it amounts to
67~meV and 120~meV per Mn interface atom from the GGA and
GGA+U calculations, respectively. In
Fig.~\ref{fig:energy} we include, for
comparison, also the curves obtained for the ideal cubic perovskite
structure (``unrelaxed''), without optimization of the atomic
positions at the interface.  The
differences between the x-values for optimal doping obtained for the two
structures are not larger
than the uncertainties of the numerical and interpolation procedures, which
indicates that the result is not sensitive to the
 change from the ideal cubic perovskite to the optimized interface
structure~\cite{note_relaxation}.
We also note that the $\it ab~initio$ optimal doping is different from
the predictions of the model with the nominal charges
for the La, Sr, Ti, and O ions, in which the  estimated 
optimal doping is around $\rm x=2/3$. 

\begin{figure}[t]
\begin{center}
\includegraphics[width=8.3cm,angle=0]{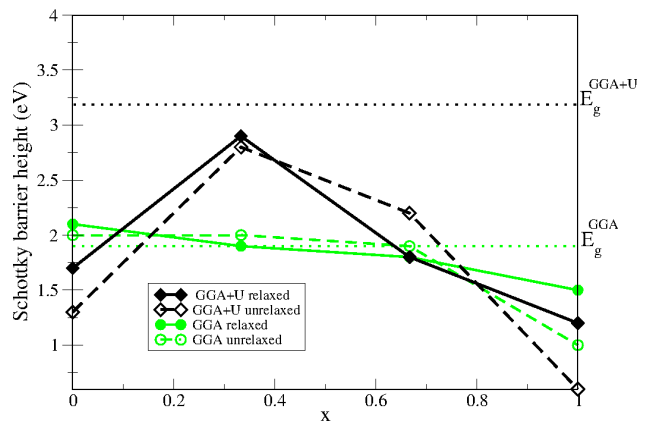}
\caption{The $p$-type Schottky barrier height as a function of doping,
  shown for the relaxed and unrelaxed cases in GGA+U and GGA. Also the
values of the calculated STO band gaps (dotted lines) are shown. $E_g^{GGA+U}$ and
$E_g^{GGA}$ stand for the STO bandgap values calculated in GGA+U and GGA, respectively.}
\label{fig:SBH}
\end{center}
\end{figure}

In Fig.~\ref{fig:SBH} we present the calculated values for the $p$-type
SBH~\cite{ZheBin10} as a function of doping for GGA+U and GGA. The
largest  SBH within the GGA+U calculations is found at $\rm
x\approx1/3$, $\it i.e.$ at about the same composition which gives
the optimal doping for ferromagnetism. It is evident that the
optimization of the atomic structure does not alter the trend. For the
change of x from 1 to 1/3 [from SrO to $\rm TiO_2$-terminated
LSMO/STO(001)], we observe a quasi linear increase of the $p$-type SBH,
in agreement with the experimental results~\cite{HikNisYaj09} and
previous theoretical findings (available for the interval $\rm
0.5<x<1$)~\cite{BurTsy10}. This trend can be expected from the consideration
of the
interface dipole, which linearly increases from  x~=~1 towards lower
x, as the electronic charge
donated by the $\rm La_{1-x}Sr_x$ to the Mn grows linearly. The change of 
trend occurs abruptly near x~=~1/3, when the GGA+U SBH starts decreasing towards the
lower value at x~=~0. By evaluating the 3d electronic charge on the
Ti and  Mn ions,
we established that this change of trend is due to a change in the
d-valence charge of the Ti ions, from $q_d\approx 0$ ($d^0$) for
$x\gtrsim1/3$ to $q_d\approx0.5$~e for x~=~0 with a related decrease
(by 0.1~e) of the Mn d electronic charge from x~=~1/3 to x~=~0. Hence,
when decreasing x from 1/3, an increasing amount of the $\rm
La_{1-x}Sr_x$ electronic dopant charge is transferred to the
neighboring Ti ions rather than to the Mn ion. 
 This charge transfer causes a dipole moment of opposite orientation,
which partially cancels the original $\rm La_{1-x}Sr_x-Mn$ dipole and
thus lowers the SBH. In fact, a
similar charging of the  interfacial Ti was previously reported
experimentally at the $\rm STO/LaMnO_3(001)$ interface~\cite{GarCezBru10}.
We note that the deviation from the nominal valence charge of the Ti
ions for $x<1/3$ also implies that the model
description~\cite{ZenGehTem07,ZheBin10} considering the Mn as the only
mixed valence ions looses its validity for $x\lesssim1/3$. This,
together with the model-inherent constraints and approximations in the
description of the valence charge for all x, clearly limits its predictive
capabilities for a quantitative assessment of the optimal doping. 

In this paper, we have calculated the optimal doping for a robust
ferromagnetism at the LSMO/STO interface by means of first-principles
DFT calculations and obtained the value
$x\approx1/3$. It corresponds to
the concentration for which the interface becomes abrupt
with the $\rm TiO_2$ termination of STO. At about the same composition,
we find that the $p$-type SBH has the maximum value.

\acknowledgments
We thank B. Davidson, A. Verna and A. Y. Petrov for useful
discussions. Calculations were performed on the IBM sp6 computers at
CINECA.



\bibliography{LSMO}

\begin{thebibliography}{35}
\expandafter\ifx\csname natexlab\endcsname\relax\def\natexlab#1{#1}\fi
\expandafter\ifx\csname bibnamefont\endcsname\relax
  \def\bibnamefont#1{#1}\fi
\expandafter\ifx\csname bibfnamefont\endcsname\relax
  \def\bibfnamefont#1{#1}\fi
\expandafter\ifx\csname citenamefont\endcsname\relax
  \def\citenamefont#1{#1}\fi
\expandafter\ifx\csname url\endcsname\relax
  \def\url#1{\texttt{#1}}\fi
\expandafter\ifx\csname urlprefix\endcsname\relax\def\urlprefix{URL }\fi
\providecommand{\bibinfo}[2]{#2}
\providecommand{\eprint}[2][]{\url{#2}}

\bibitem[{\citenamefont{Kourkoutis et~al.}(2010)\citenamefont{Kourkoutis, Song,
  Hwang, and Muller}}]{KouSonHwa10}
\bibinfo{author}{\bibfnamefont{L.~F.} \bibnamefont{Kourkoutis}},
  \bibinfo{author}{\bibfnamefont{J.~H.} \bibnamefont{Song}},
  \bibinfo{author}{\bibfnamefont{H.~Y.} \bibnamefont{Hwang}}, \bibnamefont{and}
  \bibinfo{author}{\bibfnamefont{D.~A.} \bibnamefont{Muller}},
  \bibinfo{journal}{PNAS} \textbf{\bibinfo{volume}{107}},
  \bibinfo{pages}{11682} (\bibinfo{year}{2010}).

\bibitem[{\citenamefont{Yamada et~al.}(2004)\citenamefont{Yamada, Ogawa, Ishii,
  Sato, Kawasaki, Akoh, and Tokura}}]{YamOgaIsh04}
\bibinfo{author}{\bibfnamefont{H.}~\bibnamefont{Yamada}},
  \bibinfo{author}{\bibfnamefont{Y.}~\bibnamefont{Ogawa}},
  \bibinfo{author}{\bibfnamefont{Y.}~\bibnamefont{Ishii}},
  \bibinfo{author}{\bibfnamefont{H.}~\bibnamefont{Sato}},
  \bibinfo{author}{\bibfnamefont{M.}~\bibnamefont{Kawasaki}},
  \bibinfo{author}{\bibfnamefont{H.}~\bibnamefont{Akoh}}, \bibnamefont{and}
  \bibinfo{author}{\bibfnamefont{Y.}~\bibnamefont{Tokura}},
  \bibinfo{journal}{Science} \textbf{\bibinfo{volume}{305}},
  \bibinfo{pages}{646} (\bibinfo{year}{2004}).

\bibitem[{\citenamefont{G.-X.Miao et~al.}(2011)\citenamefont{G.-X.Miao,
  M{\"{u}}nzenberg, and Moodera}}]{MiaMunMoo11}
\bibinfo{author}{\bibnamefont{G.-X.Miao}},
  \bibinfo{author}{\bibfnamefont{M.}~\bibnamefont{M{\"{u}}nzenberg}},
  \bibnamefont{and} \bibinfo{author}{\bibfnamefont{J.~S.}
  \bibnamefont{Moodera}}, \bibinfo{journal}{Rep. Prog. Phys.}
  \textbf{\bibinfo{volume}{74}}, \bibinfo{pages}{036501}
  (\bibinfo{year}{2011}).

\bibitem[{\citenamefont{Yajima et~al.}(2011)\citenamefont{Yajima, Hikita, and
  H.~Y}}]{YajHikHwa11}
\bibinfo{author}{\bibfnamefont{T.}~\bibnamefont{Yajima}},
  \bibinfo{author}{\bibfnamefont{Y.}~\bibnamefont{Hikita}}, \bibnamefont{and}
  \bibinfo{author}{\bibfnamefont{H.}~\bibnamefont{H.~Y}},
  \bibinfo{journal}{Nat. Mater.} \textbf{\bibinfo{volume}{10}},
  \bibinfo{pages}{198} (\bibinfo{year}{2011}).

\bibitem[{\citenamefont{Bibes and Barthelemy}(2007)}]{BibBer07}
\bibinfo{author}{\bibfnamefont{M.}~\bibnamefont{Bibes}} \bibnamefont{and}
  \bibinfo{author}{\bibfnamefont{A.}~\bibnamefont{Barthelemy}},
  \bibinfo{journal}{IEEE Trans. Electron Devices}
  \textbf{\bibinfo{volume}{54}}, \bibinfo{pages}{1003} (\bibinfo{year}{2007}).

\bibitem[{not({\natexlab{a}})}]{note_papers}
\eprint{M. Bowen, M. Bibes, A. Barthelemy, J. P. Contour, A. Anane, Y. 251
  Lemaitre, and A. Fert, Appl. Phys. Lett. 82, 233 (2003); R. Werner, A. Y.
  Petrov, L. A. Mino, R. Kleiner, D. Koelle, B. A. Davidson, Appl. Phys. Lett.
  98, 162505 (2011).}

\bibitem[{\citenamefont{Garcia et~al.}(2004)\citenamefont{Garcia, Bibes,
  Barth\'el\'emy, Bowen, Jacquet, Contour, and Fert}}]{GarBibBar04}
\bibinfo{author}{\bibfnamefont{V.}~\bibnamefont{Garcia}},
  \bibinfo{author}{\bibfnamefont{M.}~\bibnamefont{Bibes}},
  \bibinfo{author}{\bibfnamefont{A.}~\bibnamefont{Barth\'el\'emy}},
  \bibinfo{author}{\bibfnamefont{M.}~\bibnamefont{Bowen}},
  \bibinfo{author}{\bibfnamefont{E.}~\bibnamefont{Jacquet}},
  \bibinfo{author}{\bibfnamefont{J.-P.} \bibnamefont{Contour}},
  \bibnamefont{and} \bibinfo{author}{\bibfnamefont{A.}~\bibnamefont{Fert}},
  \bibinfo{journal}{Phys. Rev. B} \textbf{\bibinfo{volume}{69}},
  \bibinfo{pages}{052403} (\bibinfo{year}{2004}).

\bibitem[{\citenamefont{Sun et~al.}(1996)\citenamefont{Sun, Gallagher,
  Duncombe, Krusin-Elbaum, Altman, Gupta, Lu, Gong, and Xiao}}]{SunGalDun96}
\bibinfo{author}{\bibfnamefont{J.~Z.} \bibnamefont{Sun}},
  \bibinfo{author}{\bibfnamefont{W.~J.} \bibnamefont{Gallagher}},
  \bibinfo{author}{\bibfnamefont{P.~R.} \bibnamefont{Duncombe}},
  \bibinfo{author}{\bibfnamefont{L.}~\bibnamefont{Krusin-Elbaum}},
  \bibinfo{author}{\bibfnamefont{R.~A.} \bibnamefont{Altman}},
  \bibinfo{author}{\bibfnamefont{A.}~\bibnamefont{Gupta}},
  \bibinfo{author}{\bibfnamefont{Y.}~\bibnamefont{Lu}},
  \bibinfo{author}{\bibfnamefont{G.~Q.} \bibnamefont{Gong}}, \bibnamefont{and}
  \bibinfo{author}{\bibfnamefont{G.}~\bibnamefont{Xiao}},
  \bibinfo{journal}{Appl. Phys. Lett.} \textbf{\bibinfo{volume}{69}},
  \bibinfo{pages}{3266} (\bibinfo{year}{1996}).

\bibitem[{\citenamefont{Gray et~al.}(2010)\citenamefont{Gray, Papp, Balke,
  Yang, Huijben, Rotenberg, Bostwick, Ueda, Yamashita, Kobayashi
  et~al.}}]{GraPapBal10}
\bibinfo{author}{\bibfnamefont{A.~X.} \bibnamefont{Gray}},
  \bibinfo{author}{\bibfnamefont{C.}~\bibnamefont{Papp}},
  \bibinfo{author}{\bibfnamefont{B.}~\bibnamefont{Balke}},
  \bibinfo{author}{\bibfnamefont{S.-H.} \bibnamefont{Yang}},
  \bibinfo{author}{\bibfnamefont{M.}~\bibnamefont{Huijben}},
  \bibinfo{author}{\bibfnamefont{E.}~\bibnamefont{Rotenberg}},
  \bibinfo{author}{\bibfnamefont{A.}~\bibnamefont{Bostwick}},
  \bibinfo{author}{\bibfnamefont{S.}~\bibnamefont{Ueda}},
  \bibinfo{author}{\bibfnamefont{Y.}~\bibnamefont{Yamashita}},
  \bibinfo{author}{\bibfnamefont{K.}~\bibnamefont{Kobayashi}},
  \bibnamefont{et~al.}, \bibinfo{journal}{Phys. Rev. B}
  \textbf{\bibinfo{volume}{82}}, \bibinfo{pages}{205116}
  (\bibinfo{year}{2010}).

\bibitem[{\citenamefont{Herger et~al.}(2008)\citenamefont{Herger, Willmott,
  Schlep\"utz, Bj\"orck, Pauli, Martoccia, Patterson, Kumah, Clarke, Yacoby
  et~al.}}]{HerWilSch08}
\bibinfo{author}{\bibfnamefont{R.}~\bibnamefont{Herger}},
  \bibinfo{author}{\bibfnamefont{P.~R.} \bibnamefont{Willmott}},
  \bibinfo{author}{\bibfnamefont{C.~M.} \bibnamefont{Schlep\"utz}},
  \bibinfo{author}{\bibfnamefont{M.}~\bibnamefont{Bj\"orck}},
  \bibinfo{author}{\bibfnamefont{S.~A.} \bibnamefont{Pauli}},
  \bibinfo{author}{\bibfnamefont{D.}~\bibnamefont{Martoccia}},
  \bibinfo{author}{\bibfnamefont{B.~D.} \bibnamefont{Patterson}},
  \bibinfo{author}{\bibfnamefont{D.}~\bibnamefont{Kumah}},
  \bibinfo{author}{\bibfnamefont{R.}~\bibnamefont{Clarke}},
  \bibinfo{author}{\bibfnamefont{Y.}~\bibnamefont{Yacoby}},
  \bibnamefont{et~al.}, \bibinfo{journal}{Phys. Rev. B}
  \textbf{\bibinfo{volume}{77}}, \bibinfo{pages}{085401}
  (\bibinfo{year}{2008}).

\bibitem[{\citenamefont{Samet et~al.}(2003)\citenamefont{Samet, Imhoff,
  Maurice, Contour, Gloter, Manoubi, Fert, and Colliex}}]{SamImhMau03}
\bibinfo{author}{\bibfnamefont{L.}~\bibnamefont{Samet}},
  \bibinfo{author}{\bibfnamefont{D.}~\bibnamefont{Imhoff}},
  \bibinfo{author}{\bibfnamefont{J.-L.} \bibnamefont{Maurice}},
  \bibinfo{author}{\bibfnamefont{J.-P.} \bibnamefont{Contour}},
  \bibinfo{author}{\bibfnamefont{A.}~\bibnamefont{Gloter}},
  \bibinfo{author}{\bibfnamefont{T.}~\bibnamefont{Manoubi}},
  \bibinfo{author}{\bibfnamefont{A.}~\bibnamefont{Fert}}, \bibnamefont{and}
  \bibinfo{author}{\bibfnamefont{C.}~\bibnamefont{Colliex}},
  \bibinfo{journal}{Eur. Phys. J. B} \textbf{\bibinfo{volume}{34}},
  \bibinfo{pages}{179} (\bibinfo{year}{2003}).

\bibitem[{\citenamefont{Pailloux et~al.}(2002)\citenamefont{Pailloux, Imhoff,
  Sikora, Barthelemy, Maurice, Contour, Colliex, and Fert}}]{PaiImhSik02}
\bibinfo{author}{\bibfnamefont{F.}~\bibnamefont{Pailloux}},
  \bibinfo{author}{\bibfnamefont{D.}~\bibnamefont{Imhoff}},
  \bibinfo{author}{\bibfnamefont{T.}~\bibnamefont{Sikora}},
  \bibinfo{author}{\bibfnamefont{A.}~\bibnamefont{Barthelemy}},
  \bibinfo{author}{\bibfnamefont{I.~L.} \bibnamefont{Maurice}},
  \bibinfo{author}{\bibfnamefont{J.~P.} \bibnamefont{Contour}},
  \bibinfo{author}{\bibfnamefont{C.}~\bibnamefont{Colliex}}, \bibnamefont{and}
  \bibinfo{author}{\bibfnamefont{A.}~\bibnamefont{Fert}},
  \bibinfo{journal}{Phys.\ Rev.\ B} \textbf{\bibinfo{volume}{66}},
  \bibinfo{pages}{014417} (\bibinfo{year}{2002}).

\bibitem[{\citenamefont{Verna et~al.}(2012)\citenamefont{Verna, Davidson,
  Mirone, and Nannarone}}]{VerDavMir12}
\bibinfo{author}{\bibfnamefont{A.}~\bibnamefont{Verna}},
  \bibinfo{author}{\bibfnamefont{B.~A.} \bibnamefont{Davidson}},
  \bibinfo{author}{\bibfnamefont{A.}~\bibnamefont{Mirone}}, \bibnamefont{and}
  \bibinfo{author}{\bibfnamefont{S.}~\bibnamefont{Nannarone}},
  \bibinfo{journal}{Eur. Phys. J. Spec. Top.} \textbf{\bibinfo{volume}{208}},
  \bibinfo{pages}{165} (\bibinfo{year}{2012}).

\bibitem[{\citenamefont{Izumi et~al.}(2001)\citenamefont{Izumi, Ogimoto,
  Okimoto, Manako, Ahmet, Nakajima, Chikyow, Kawasaki, and
  Tokura}}]{IzuOgiOki01}
\bibinfo{author}{\bibfnamefont{M.}~\bibnamefont{Izumi}},
  \bibinfo{author}{\bibfnamefont{Y.}~\bibnamefont{Ogimoto}},
  \bibinfo{author}{\bibfnamefont{Y.}~\bibnamefont{Okimoto}},
  \bibinfo{author}{\bibfnamefont{T.}~\bibnamefont{Manako}},
  \bibinfo{author}{\bibfnamefont{P.}~\bibnamefont{Ahmet}},
  \bibinfo{author}{\bibfnamefont{K.}~\bibnamefont{Nakajima}},
  \bibinfo{author}{\bibfnamefont{T.}~\bibnamefont{Chikyow}},
  \bibinfo{author}{\bibfnamefont{M.}~\bibnamefont{Kawasaki}}, \bibnamefont{and}
  \bibinfo{author}{\bibfnamefont{Y.}~\bibnamefont{Tokura}},
  \bibinfo{journal}{Phys. Rev. B} \textbf{\bibinfo{volume}{64}},
  \bibinfo{pages}{064429} (\bibinfo{year}{2001}).

\bibitem[{\citenamefont{Ogimoto et~al.}(2003)\citenamefont{Ogimoto, Izumi,
  Sawa, Manako, Sato, Akoh, Kawasaki, and Tokura}}]{OgiIzuSaw03}
\bibinfo{author}{\bibfnamefont{Y.}~\bibnamefont{Ogimoto}},
  \bibinfo{author}{\bibfnamefont{M.}~\bibnamefont{Izumi}},
  \bibinfo{author}{\bibfnamefont{A.}~\bibnamefont{Sawa}},
  \bibinfo{author}{\bibfnamefont{T.}~\bibnamefont{Manako}},
  \bibinfo{author}{\bibfnamefont{H.}~\bibnamefont{Sato}},
  \bibinfo{author}{\bibfnamefont{H.}~\bibnamefont{Akoh}},
  \bibinfo{author}{\bibfnamefont{M.}~\bibnamefont{Kawasaki}}, \bibnamefont{and}
  \bibinfo{author}{\bibfnamefont{Y.}~\bibnamefont{Tokura}},
  \bibinfo{journal}{Jpn. J. Appl. Phys.} \textbf{\bibinfo{volume}{42}},
  \bibinfo{pages}{L369} (\bibinfo{year}{2003}).

\bibitem[{\citenamefont{Ishii et~al.}(2006)\citenamefont{Ishii, Yamada, Sato,
  Akoh, Ogawa, Kawasaki, and Tokura}}]{IshYamSat06}
\bibinfo{author}{\bibfnamefont{Y.}~\bibnamefont{Ishii}},
  \bibinfo{author}{\bibfnamefont{H.}~\bibnamefont{Yamada}},
  \bibinfo{author}{\bibfnamefont{H.}~\bibnamefont{Sato}},
  \bibinfo{author}{\bibfnamefont{H.}~\bibnamefont{Akoh}},
  \bibinfo{author}{\bibfnamefont{Y.}~\bibnamefont{Ogawa}},
  \bibinfo{author}{\bibfnamefont{M.}~\bibnamefont{Kawasaki}}, \bibnamefont{and}
  \bibinfo{author}{\bibfnamefont{Y.}~\bibnamefont{Tokura}},
  \bibinfo{journal}{App. Phys. Lett.} \textbf{\bibinfo{volume}{89}},
  \bibinfo{eid}{042509} (\bibinfo{year}{2006}).

\bibitem[{\citenamefont{Boschker et~al.}(2012)\citenamefont{Boschker, Verbeeck,
  Egoavil, Bals, van Tendeloo, Huijben, Houwman, Koster, Blank, and
  Rijnders}}]{BosVerEgo12}
\bibinfo{author}{\bibfnamefont{H.}~\bibnamefont{Boschker}},
  \bibinfo{author}{\bibfnamefont{J.}~\bibnamefont{Verbeeck}},
  \bibinfo{author}{\bibfnamefont{R.}~\bibnamefont{Egoavil}},
  \bibinfo{author}{\bibfnamefont{S.}~\bibnamefont{Bals}},
  \bibinfo{author}{\bibfnamefont{G.}~\bibnamefont{van Tendeloo}},
  \bibinfo{author}{\bibfnamefont{M.}~\bibnamefont{Huijben}},
  \bibinfo{author}{\bibfnamefont{E.~P.} \bibnamefont{Houwman}},
  \bibinfo{author}{\bibfnamefont{G.}~\bibnamefont{Koster}},
  \bibinfo{author}{\bibfnamefont{D.~H.~A.} \bibnamefont{Blank}},
  \bibnamefont{and} \bibinfo{author}{\bibfnamefont{G.}~\bibnamefont{Rijnders}},
  \bibinfo{journal}{Advanced Functional Materials}
  \textbf{\bibinfo{volume}{22}}, \bibinfo{pages}{2235} (\bibinfo{year}{2012}).

\bibitem[{\citenamefont{Zenia et~al.}(2007)\citenamefont{Zenia, Gehring, and
  Temmerman}}]{ZenGehTem07}
\bibinfo{author}{\bibfnamefont{H.}~\bibnamefont{Zenia}},
  \bibinfo{author}{\bibfnamefont{G.~A.} \bibnamefont{Gehring}},
  \bibnamefont{and} \bibinfo{author}{\bibfnamefont{W.~M.}
  \bibnamefont{Temmerman}}, \bibinfo{journal}{New J. Phys.}
  \textbf{\bibinfo{volume}{9}}, \bibinfo{pages}{105} (\bibinfo{year}{2007}).

\bibitem[{\citenamefont{Zheng and Binggeli}(2010)}]{ZheBin10}
\bibinfo{author}{\bibfnamefont{B.}~\bibnamefont{Zheng}} \bibnamefont{and}
  \bibinfo{author}{\bibfnamefont{N.}~\bibnamefont{Binggeli}},
  \bibinfo{journal}{Phys. Rev. B} \textbf{\bibinfo{volume}{82}},
  \bibinfo{pages}{245311} (\bibinfo{year}{2010}).

\bibitem[{\citenamefont{Burton and Tsymbal}(2010)}]{BurTsy10}
\bibinfo{author}{\bibfnamefont{J.~D.} \bibnamefont{Burton}} \bibnamefont{and}
  \bibinfo{author}{\bibfnamefont{E.~Y.} \bibnamefont{Tsymbal}},
  \bibinfo{journal}{Phys. Rev. B} \textbf{\bibinfo{volume}{82}},
  \bibinfo{pages}{161407R} (\bibinfo{year}{2010}).

\bibitem[{not({\natexlab{b}})}]{note_model}
\eprint{Besides nominal charges, this model assumes charge neutrality and that
  the Mn charge in the junction is different from that in bulk LSMO only in the
  $\rm MnO_2$ monolayer closest to the STO. The optimal doping for the
  interface $\rm MnO_2$ layer in this model is postulated to be the one which
  yields the same charge on the Mn ion as in the bulk LSMO ($\rm Mn^{+10/3}$).}

\bibitem[{\citenamefont{Bowen et~al.}(2006)\citenamefont{Bowen, Barth\'el\'emy,
  Bellini, Bibes, Seneor, Jacquet, Contour, and Dederichs}}]{BowBarBel06}
\bibinfo{author}{\bibfnamefont{M.}~\bibnamefont{Bowen}},
  \bibinfo{author}{\bibfnamefont{A.}~\bibnamefont{Barth\'el\'emy}},
  \bibinfo{author}{\bibfnamefont{V.}~\bibnamefont{Bellini}},
  \bibinfo{author}{\bibfnamefont{M.}~\bibnamefont{Bibes}},
  \bibinfo{author}{\bibfnamefont{P.}~\bibnamefont{Seneor}},
  \bibinfo{author}{\bibfnamefont{E.}~\bibnamefont{Jacquet}},
  \bibinfo{author}{\bibfnamefont{J.-P.} \bibnamefont{Contour}},
  \bibnamefont{and} \bibinfo{author}{\bibfnamefont{P.~H.}
  \bibnamefont{Dederichs}}, \bibinfo{journal}{Phys. Rev. B}
  \textbf{\bibinfo{volume}{73}}, \bibinfo{pages}{140408}
  (\bibinfo{year}{2006}).

\bibitem[{\citenamefont{Giannozzi et~al.}(2009)\citenamefont{Giannozzi, Baroni,
  Bonini, Calandra, Car, Cavazzoni, Ceresoli, Chiarotti, Cococcioni, Dabo
  et~al.}}]{GiaBarBon09}
\bibinfo{author}{\bibfnamefont{P.}~\bibnamefont{Giannozzi}},
  \bibinfo{author}{\bibfnamefont{S.}~\bibnamefont{Baroni}},
  \bibinfo{author}{\bibfnamefont{N.}~\bibnamefont{Bonini}},
  \bibinfo{author}{\bibfnamefont{M.}~\bibnamefont{Calandra}},
  \bibinfo{author}{\bibfnamefont{R.}~\bibnamefont{Car}},
  \bibinfo{author}{\bibfnamefont{C.}~\bibnamefont{Cavazzoni}},
  \bibinfo{author}{\bibfnamefont{D.}~\bibnamefont{Ceresoli}},
  \bibinfo{author}{\bibfnamefont{G.~L.} \bibnamefont{Chiarotti}},
  \bibinfo{author}{\bibfnamefont{M.}~\bibnamefont{Cococcioni}},
  \bibinfo{author}{\bibfnamefont{I.}~\bibnamefont{Dabo}}, \bibnamefont{et~al.},
  \bibinfo{journal}{J. Phys.: Condens. Matter} \textbf{\bibinfo{volume}{21}},
  \bibinfo{pages}{395502} (\bibinfo{year}{2009}).

\bibitem[{\citenamefont{Vanderbilt}(1990)}]{Van90}
\bibinfo{author}{\bibfnamefont{D.}~\bibnamefont{Vanderbilt}},
  \bibinfo{journal}{Phys.\ Rev.\ B} \textbf{\bibinfo{volume}{41}},
  \bibinfo{pages}{7892} (\bibinfo{year}{1990}).

\bibitem[{pse()}]{pseudo}
\eprint{The following reference atomic configurations were used for the
  generation of the pseudopotentials: 3d$^5$4s$^2$4p$^0$ for Mn,
  5s$^2$5p$^6$5d$^1$6s$^{1.5}$6p$^{0.5}$ for La, 4s$^2$4p$^6$4d$^1$5s$^1$5p$^0$
  for Sr, 3s$^2$3p$^6$3d$^1$4s$^2$ for Ti, and 2s$^2$2p$^4$ for O. The
  nonlinear core-correction to the exchange-correlation potential was included
  for Mn, La, and Sr.}

\bibitem[{\citenamefont{Perdew et~al.}(1996)\citenamefont{Perdew, Burke, and
  Ernzerhof}}]{PerBurErn96}
\bibinfo{author}{\bibfnamefont{J.~P.} \bibnamefont{Perdew}},
  \bibinfo{author}{\bibfnamefont{K.}~\bibnamefont{Burke}}, \bibnamefont{and}
  \bibinfo{author}{\bibfnamefont{M.}~\bibnamefont{Ernzerhof}},
  \bibinfo{journal}{Phys.\ Rev.\ Lett.} \textbf{\bibinfo{volume}{77}},
  \bibinfo{pages}{3865} (\bibinfo{year}{1996}).

\bibitem[{\citenamefont{Anisimov et~al.}(1997)\citenamefont{Anisimov,
  Aruasetiawan, and Lichtenstein}}]{AniAruLic97}
\bibinfo{author}{\bibfnamefont{V.~I.} \bibnamefont{Anisimov}},
  \bibinfo{author}{\bibfnamefont{F.}~\bibnamefont{Aruasetiawan}},
  \bibnamefont{and} \bibinfo{author}{\bibfnamefont{A.~I.}
  \bibnamefont{Lichtenstein}}, \bibinfo{journal}{J.\ Phys.\ :Condens.\ Matter}
  \textbf{\bibinfo{volume}{9}}, \bibinfo{pages}{767} (\bibinfo{year}{1997}).

\bibitem[{\citenamefont{Zheng and Binggeli}(2009)}]{ZheBin09}
\bibinfo{author}{\bibfnamefont{B.}~\bibnamefont{Zheng}} \bibnamefont{and}
  \bibinfo{author}{\bibfnamefont{N.}~\bibnamefont{Binggeli}},
  \bibinfo{journal}{J. Phys.: Condens. Matter} \textbf{\bibinfo{volume}{21}},
  \bibinfo{pages}{115602} (\bibinfo{year}{2009}).

\bibitem[{\citenamefont{Tsui et~al.}(2000)\citenamefont{Tsui, Smoak, Nath, and
  Eom}}]{TsuSmoNat00}
\bibinfo{author}{\bibfnamefont{F.}~\bibnamefont{Tsui}},
  \bibinfo{author}{\bibfnamefont{M.~C.} \bibnamefont{Smoak}},
  \bibinfo{author}{\bibfnamefont{T.~K.} \bibnamefont{Nath}}, \bibnamefont{and}
  \bibinfo{author}{\bibfnamefont{C.~B.} \bibnamefont{Eom}},
  \bibinfo{journal}{Appl.\ Phys.\ Lett.} \textbf{\bibinfo{volume}{76}},
  \bibinfo{pages}{2421} (\bibinfo{year}{2000}).

\bibitem[{\citenamefont{Woicik et~al.}(2006)\citenamefont{Woicik, Li, Zschack,
  Karapetrova, Ryan, Ashman, and Hellberg}}]{WoiLiZsc06}
\bibinfo{author}{\bibfnamefont{J.~C.} \bibnamefont{Woicik}},
  \bibinfo{author}{\bibfnamefont{H.}~\bibnamefont{Li}},
  \bibinfo{author}{\bibfnamefont{P.}~\bibnamefont{Zschack}},
  \bibinfo{author}{\bibfnamefont{E.}~\bibnamefont{Karapetrova}},
  \bibinfo{author}{\bibfnamefont{P.}~\bibnamefont{Ryan}},
  \bibinfo{author}{\bibfnamefont{C.~R.} \bibnamefont{Ashman}},
  \bibnamefont{and} \bibinfo{author}{\bibfnamefont{C.~S.}
  \bibnamefont{Hellberg}}, \bibinfo{journal}{Phys.\ Rev.\ B}
  \textbf{\bibinfo{volume}{73}}, \bibinfo{pages}{024112}
  (\bibinfo{year}{2006}).

\bibitem[{\citenamefont{Geng and Zhang}(2006)}]{GenZha06}
\bibinfo{author}{\bibfnamefont{T.}~\bibnamefont{Geng}} \bibnamefont{and}
  \bibinfo{author}{\bibfnamefont{N.}~\bibnamefont{Zhang}},
  \bibinfo{journal}{Phys.\ Lett.\ A} \textbf{\bibinfo{volume}{351}},
  \bibinfo{pages}{314} (\bibinfo{year}{2006}).

\bibitem[{\citenamefont{van Benthem et~al.}(2001)\citenamefont{van Benthem,
  Elsasser, and French}}]{BenElsFre01}
\bibinfo{author}{\bibfnamefont{K.}~\bibnamefont{van Benthem}},
  \bibinfo{author}{\bibfnamefont{C.}~\bibnamefont{Elsasser}}, \bibnamefont{and}
  \bibinfo{author}{\bibfnamefont{R.~H.} \bibnamefont{French}},
  \bibinfo{journal}{J.\ Appl.\ Phys.} \textbf{\bibinfo{volume}{90}},
  \bibinfo{pages}{6156} (\bibinfo{year}{2001}).

\bibitem[{not({\natexlab{c}})}]{note_relaxation}
\eprint{This might seem surprising, as the relaxations of the individual atoms
  at the interface are significant (displacements around $\sim0.1$~\AA~
  perpendicular to the interface~\cite{ZheBin10,BurTsy10}), but the relaxations
  are similar for the FM and AFM interface spin alignments.}

\bibitem[{\citenamefont{Hikita et~al.}(2009)\citenamefont{Hikita, Nishikawa,
  Yajima, and Hwang}}]{HikNisYaj09}
\bibinfo{author}{\bibfnamefont{Y.}~\bibnamefont{Hikita}},
  \bibinfo{author}{\bibfnamefont{M.}~\bibnamefont{Nishikawa}},
  \bibinfo{author}{\bibfnamefont{T.}~\bibnamefont{Yajima}}, \bibnamefont{and}
  \bibinfo{author}{\bibfnamefont{H.~Y.} \bibnamefont{Hwang}},
  \bibinfo{journal}{Phys. Rev. B} \textbf{\bibinfo{volume}{79}},
  \bibinfo{pages}{073101} (\bibinfo{year}{2009}).

\bibitem[{\citenamefont{Garcia-Barriocanal
  et~al.}(2010)\citenamefont{Garcia-Barriocanal, Cezar, Bruno, Thakur, Brookes,
  Utfeld, Rivera-Calzada, Giblin, Taylor, Duffy et~al.}}]{GarCezBru10}
\bibinfo{author}{\bibfnamefont{J.}~\bibnamefont{Garcia-Barriocanal}},
  \bibinfo{author}{\bibfnamefont{J.}~\bibnamefont{Cezar}},
  \bibinfo{author}{\bibfnamefont{F.}~\bibnamefont{Bruno}},
  \bibinfo{author}{\bibfnamefont{P.}~\bibnamefont{Thakur}},
  \bibinfo{author}{\bibfnamefont{N.}~\bibnamefont{Brookes}},
  \bibinfo{author}{\bibfnamefont{C.}~\bibnamefont{Utfeld}},
  \bibinfo{author}{\bibfnamefont{A.}~\bibnamefont{Rivera-Calzada}},
  \bibinfo{author}{\bibfnamefont{S.}~\bibnamefont{Giblin}},
  \bibinfo{author}{\bibfnamefont{J.}~\bibnamefont{Taylor}},
  \bibinfo{author}{\bibfnamefont{J.}~\bibnamefont{Duffy}},
  \bibnamefont{et~al.}, \bibinfo{journal}{Nat. Commun.}
  \textbf{\bibinfo{volume}{1}}, \bibinfo{pages}{82} (\bibinfo{year}{2010}).

\end{thebibliography}
\end{document}